%Paper: hep-ph/9307280
%From: LADINSKY@MSUPA.PA.MSU.EDU
%Date: Thu, 15 Jul 1993 17:13:36 -0400 (EDT)

%%%%%%%%%%%%%%%%%%%%%%%%%%%%%%%%%%%%%%%%%%%%%%%%%%%%%%%%%%%%%%%%%%%%
%%%    This paper uses the TeXsis macro with the WorldSci.txs    %%%
%%%           style file (or the mtexsis file on hepph).         %%%
%%%%%%%%%%%%%%%%%%%%%%%%%%%%%%%%%%%%%%%%%%%%%%%%%%%%%%%%%%%%%%%%%%%%
\texsis
\WorldScientific
%========================DEFINITIONS===================================
\def\Ztt{\hbox{$Z$-$t$-$\bar t$}}

\def\A{{\cal A}}

\def\Lag{{\cal L}}
\def\del{\partial}
\def\journal#1&#2(#3)#4{
{\unskip,~\sl #1\unskip~\bf\ignorespaces #2\unskip~\rm (19#3) #4}}
\def\jour#1&#2(#3)#4{
{\unskip ~\sl #1\unskip~\bf\ignorespaces #2\unskip~\rm (19#3) #4}}
\def\eett{{$e^-e^+\to t\bar{t}$}}
\def\uline#1{{$\underline{\hbox{#1}}$}}
%=========================Title Page===============================
%\titlepage
{\tenpoint 15 July 1993 \hfill MSUTH 93/09}
\title{ Probing Form Factors in Top Quark
Pair Production at $e^-e^+$ Colliders }
\author
G.~A. Ladinsky \ \ and \ \ \uline{C.--P. Yuan}
{\it Michigan State University, Department of Physics and Astronomy}
{\it East Lansing, MI 48824}
\endauthor

\abstract
{\tenpoint
We describe how to probe new physics through large CP violation
effects and non--standard~\Ztt\ couplings via the scattering
process~\eett.
}
\endabstract
\footnote{}{\ninerm\baselineskip=11pt Talk presented at the Workshop on
Physics and Experiments with Linear $e^+e^-$ Colliders, Waikoloa,
Hawaii, 26-30 April 1993.}
%\endtitlepage
%
%===================Reference Listing======================================
%
\referencelist
\reference{tpol}G.L.~Kane, G.A.~Ladinsky and C.--P.~Yuan
\journal Phys.~Rev. &D45 (92) 124
\endreference
\reference{peskin}
C.R.~Schmidt and M.E.~Peskin \journal Phys.~Rev.~Lett.&69 (92) 410
\endreference
\reference{weinberg}
S.~Weinberg \journal Phys.~Rev.~Lett.&63 (89) 2333;
\jour Phys.~Rev.&D42 (90) 860
\endreference
\reference{minuit}MINUIT, Application Software Group, Computing and Networks
Division, CERN, Geneva, Switzerland
\endreference
\reference{nlc}
We use NLC to represent a generic $e^-e^+$ supercollider;
for a list of proposed colliders see, e.g., R.~Settles,
in the {\it Proceedings of the
27th Rencontre de Moriond: Electroweak Interactions and Unified
Theories}, Les Arcs, France, Mar 15-22, 1992
\endreference
\reference{gal}
G.~A. Ladinsky and C.--P. Yuan, in preparation
\endreference
\endreferencelist
%
%=========================Body of paper====================================
%

Large CP violating effects are required to have the cosmological
baryon asymmetry produced at the weak phase transition.
If such effects exist, can they be probed at electron
colliders?
How well can the form factors of the \Ztt\ coupling be measured in both the
$e^-e^+\to t\bar{t}\to bl^+\nu_l\bar{b}l^-\bar{\nu_l}$
and
$e^-e^+\to t\bar{t}\to bl^+\nu_l\bar{b}qq'$
decay modes?
What are the preferred energies and luminosities for detecting large
CP violating effects and measuring the form factors governing the \Ztt\
interaction?
We addressed these issues at this conference.

The most general
form factors for the coupling
of $t$ and $\bar t$ with either of the vector bosons $\gamma$ or $Z$
can be derived from the lagrangian\cite{tpol}
$$\eqalign{
\Lag_{int}&= g\bigg\lbrack Z_\mu\bar{t}\gamma^\mu(F_1^{Z(L)} P_-
+F_1^{Z(R)} P_+)t
-{1\over v}\del_\nu Z_\mu\bar{t}\sigma^{\mu\nu}
(F_2^{Z(L)}P_-+F_2^{Z(R)}P_+)t	\cr
+&\del^\mu Z_\mu \bar{t}(F_3^{Z(L)}P_-+F_3^{Z(R)}P_+)t
+A_\mu\bar{t}\gamma^\mu(F_1^{\gamma (L)} P_-+F_1^{\gamma (R)} P_+)t
\cr
-&{1\over v} \del_\nu A_\mu\bar{t}\sigma^{\mu\nu}
(F_2^{\gamma (L)}P_-+F_2^{\gamma (R)}P_+)t
+\del^\mu A_\mu \bar{t}(F_3^{\gamma (L)}P_-+F_3^{\gamma (R)}P_+)t\bigg\rbrack
\cr}
\EQN intlag
$$
where $P_\pm ={1\over 2}(1\pm \gamma_5)$,
$i\sigma^{\mu\nu}=-{1\over 2}[\gamma^\mu,\gamma^\nu]$,
and $v\sim 250\,$GeV is the vacuum expectation value.
Since we ignore the masses of the incoming electron and positron, the
$F_3$ form factors in \Eq{intlag} do not contribute.

If the form factor $D$, defined as
$
D_{[\gamma,Z]}\equiv{2\over v}(F_2^{[\gamma,Z](L)}-F_2^{[\gamma,Z](R)}),
%\EQN dformula
$
is not zero, the theory is CP
violating.\cite{tpol}
Specifically, the difference
between the $RR$ and $LL$ cross sections is linearly dependent on $D$ through
$$
(RR-LL)\propto \Re\Bigg\lbrack\bigg\lparen {A_Z\over s-m_Z^2}
		+{A_\gamma\over s}\bigg\rparen D_Z^*\Bigg\rbrack,
\EQN cpviol
$$
where
$$
A_{[\gamma,Z]}=F_1^{[\gamma,Z](L)} +F_1^{[\gamma,Z](R)}-
{2m_t\over v}(F_2^{[\gamma,Z](L)}+F_2^{[\gamma,Z](R)}),
\EQN aformula
$$
the ``$*$'' indicates complex conjugation, and $\Re$ takes the real part
of its argument.
(In this work the photon is assumed to preserve its Standard Model
(SM) behavior, namely, $D_\gamma=0$.)
$LL$, $RR$, $LR$, $RL$ respectively denote the production rates for
$t_L \bar t_L$, $t_R \bar t_R$, $t_L \bar t_R$, $t_R \bar t_ L$
events, where $L$ labels a left--handed
helicity and $R$ labels a right--handed helicity.
{}From \Eq{cpviol} one can construct a quantity sensitive to CP violation,
$
\A_1\equiv {LL-RR\over LL+RR},
%\EQN eseven
$
to measure the real part of $D$.\cite{peskin}
The imaginary part of $D$ can be examined by studying the transverse
polarization of the top quark pairs perpendicular to the scatter
plane.\cite{tpol}
The advantage of determining the polarization states
of the $t\bar{t}$ pairs is that one can
measure $\A_1$ instead of $\A_2\equiv{LL-RR\over LL+RR+LR+RL}$ (which may lose
some of its effectiveness due to the dilution acquired by having a larger
denominator from the $RL$ and $LR$ events) when studying
CP violation.

To observe CP violation with
 $\A_1\sim 10^{-2}$, as in Weinberg's model,\cite{weinberg}
  in top quarks of $m_t=140\,$GeV, requires a luminosity of about
 $3\times 10^4\,\hbox{fb}^{-1}$ at $\sqrt{s}=500\,$GeV
(i.e., about $2\times 10^7$ $t\bar t$ pairs are required).

Below we present bounds that represent a $90\%$ and $68\%$
confidence level
on the range for determining the form factors of the \Ztt\ interaction.
(The $90\%$ ($68\%$) confidence level roughly represents a 3.3 (1)
standard deviation error.)
These bounds were obtained using MINUIT\cite{minuit}
to fit the polar angle distribution
of the top quark as generated for the NLC ($\sqrt{s}$=500 GeV, with
$50 \, {\rm fb}^{-1}$ integrated luminosity) with no constraints on the
kinematics.\cite{nlc}
We expect 30,000 $t\bar{t}$ events for $m_t=140\,$GeV,
but if we focus on the mode where the two $W$'s decay leptonically to
$\mu^\pm$ or $e^\pm$, there is a branching ratio reduction to 1,500 events.
These 1,500 events were collected into twenty bins for the fit.
Allowing only one form factor to vary at a time, the results
%presented in \Tbl{ttwo}
indicate that within the $90\%$ ($68\%$)
confidence limit, it should be
possible to find $F_1^{(L)}$ to within about $20\%$ ($6\%$), while
$F_1^{(R)}$ can only be known to within roughly $40\%$ ($13\%$).
The $F_2$ form factors are zero at the Born level
in the SM, and the fit indicates that their
values can be known to within about $0.0^{+0.148}_{-0.043}$
($0.0^{+0.022}_{-0.015}$) at a confidence level of $90\%$ ($68\%$).

Using our knowledge of
the polarization behavior of the top quark
to untangle some of the form factors before
comparing them with data, it may be possible to improve these bounds.
Performing the same procedures as we did with the unpolarized cross section,
we bin 750 events from the unconstrained $LR$ sample and use MINUIT to fit for
$F_1^{Z(L)}$ and find $0.395\pm 0.014$ at the $68\%$ confidence level,
which is roughly $2\%$ better
than the fit from the unpolarized distribution.
Further improvement in the form factor determination can come from the
increased statistics (by about a factor of 7)
obtained by including the events from
\hbox{$e^-e^+\to t\bar{t}\to l+\hbox{jets}$} in the analysis.
For the $l+\hbox{jets}$ mode, where the branching ratios
take $9,000$ unpolarized events from the original $30,000$,
the results indicate that within the $90\%$ ($68\%$)
confidence limit, it should be
possible to find $F_1^{(L)}$ to within about $8\%$ ($3\%$), while
$F_1^{(R)}$ can be known to within roughly $18\%$ ($5\%$).
In this case the $F_2$ form factors
can be known to within about $\pm 0.008$
($\pm 0.03$) at a confidence level of $90\%$ ($68\%$).

A $1\,$TeV machine can do better than a $500\,$GeV machine in determining
$F_1^{(L,R)}$ because the relative sizes of the $RR$ and $LL$ rates become
smaller compared to the total event rate, however, for the same reason a
$1\,$TeV machine makes the CP asymmetry measurement more difficult.

In reality, the initial state electron or positron at high energies
will radiate photons along the beam axis either due to initial state
radiation (ISR) or beamstrahlung.
Such radiation tends to move along the beam direction
such that neither the center--of--mass energy nor the boost
of the $t \bar t$ pair is known.
Despite this, the effects of ISR are such that
half of the time the center--of--mass energy of the $t \bar t$ pair is
very close to the beam energy.
Furthermore, it is possible to design experiments such that the
beamstrahlung is minimized, so we only consider the ISR
effect.  We checked that it is possible to solve
for the momentum of the top quark in both the
$e^-e^+\to t\bar{t}\to bl^+\nu_l\bar{b}qq'$ and
$e^-e^+\to t\bar{t}\to bl^+\nu_l\bar{b}l^-\bar{\nu_l}$
decay modes.\cite{gal}

Even including the effects of ISR,
the kinematics of the top quark can be reconstructed in the dilepton modes.
The advantage of a high energy machine in this regard is that the top
quark is highly boosted, making it easy to select the correct $b$ quark
(not the $\bar{b}$) associated with the $W^+$ to reconstruct the top
momentum.  We have found that the probability of choosing the wrong
$b$ quark is less than a percent at $\sqrt{s}=500\,$GeV and $m_t=140\,$GeV.
Determining the top quark kinematics when one $W$ boson decays hadronically
while the other $W$ boson decays leptonically is even less complicated.

We thank D.~Burke, D.~Chang, K.~Fujii, G.~Kane
and M.~Peskin for useful discussions.  This work was funded in part by
TNRLC grant \#RGFY9240.

%
% ===============Now dump the captions and references==================
%
\nosechead{References}

\ListReferences
\vfill\eject
\end